\documentclass[prl,twocolumn,epsf,psfig]{revtex4}
\usepackage{graphicx}
\DeclareGraphicsExtensions{.pdf,.png,.gif,.jpg}
\usepackage{float}
\usepackage{subfig}
\usepackage{epsfig}
\usepackage{wrapfig}
\usepackage{bbold}

\begin{document}

\title{Molecular Self-Organization of Three-Component Lipid Membranes}

\author{Austin Osby and Theja N. De Silva}
\affiliation{Department of Chemistry and Physics,
Augusta University, Augusta, Georgia 30912, USA.}

\begin{abstract}
By constructing a Landau-like energy functional, we investigate the molecular organization of a three-component mixture in cell membranes. In the strongly interacting limit, we model the interaction between molecules using pseudospin variables and convert them into non-interacting variables using a mean-field theory. Next, we construct the two-order parameter Landau-type energy functional through the Helmholtz free energy. By analyzing the Landau free energy, we map out the phase diagram focusing on homogeneous and various phase separated states on the cell membrane.
\end{abstract}

\maketitle

\section{I. Introduction}

Self-organization of matter due to the collective behavior of constituent elements is a thermodynamic process that can be seen in variety of dynamical systems~\cite{Newp1}, including biological systems~\cite{Newp2}. The cell membrane is one of the classic examples of biological systems that show self-assembly of molecules. The cell, which is the fundamental building block of all living matter, consists of cytoplasm enclosed within a membrane where the cell membrane is composed of lipids and protein molecules. The cell membrane controls the vital role of providing a barrier of separation from the external environment; as well as the unique ability to selectively allow permeability of molecules to and from the cell through active and passive transport~\cite{mREV}. In addition, the cell membrane allows flow of information between cellular contents and external environment. In order to support these cellular functions, the cell membrane must be mechanically stable yet malleable enough to allow cell growth, cell division, and shape change for change dynamics. As stated, lipids are the basic building block of membranes. Most of the lipids  in cell membranes share the same the same basic molecular structure by self-assembling in to certain structures. It is the purpose of this work to study the molecular structure of cell membranes for a three-component mixture. As our motivation is to study the possible molecular arrangements in the cell, we do not consider bi-layer formation or lipid raft formation due to the polar headgroup within lipids.

In a multi-component molecular system, the chemical interactions between different species are responsible for the self-organization of the molecules~\cite{Ip1, Ip2, Ip3, Ip4, Ip5}. At higher temperatures, the multi components within the system tend to form a homogeneous phase due to the entropy domination. However, at lower temperatures, the homogenous mixed phase is no longer thermodynamically stable and the different species tend to separate into domains at a certain critical temperature. The critical temperature and the number of fully or partially separated domains depend on the interaction between species. It is believed that the raft formation is due to the inhomogeneous distribution of lipids resulting from the phase separation of saturated lipids, unsaturated lipids, and cholesterol~\cite{Ep1,Ep2, Ep3}. Theoretically, the phase separation has been extensively studied for binary mixtures on cell membranes~\cite{Tp1, Tp2, Tp3, Tp4, Tp5,Tp6, Tp7, Tp8, Tp9, Tp10, Tp11, Tp12, Tp13}, however, ternary systems have not been studies as extensively~\cite{tt1}. Here we study the lipid organization in a ternary mixture, such as saturated lipids, unsaturated lipids, and cholesterol on cell membranes using a thermodynamical approach. We consider the strong interaction limit where the kinetic energy of the molecules can be neglected relative to the interaction energies. As a result, the molecules tend repel each other to minimize the interaction energies. For this case, the occupation of molecules at a given point in space can be represented by a pseudospin. Thus, the phase separation or the co-existence of the different molecules due to the competition between ground state energy of pseudospins and entropy is determined by the collective behaviour of the molecules. By analyzing the Helmholtz free energy and then deriving a Landau type energy functional, we construct the phase diagram in temperature-interaction parameter space. We find that the phase separation and the co-existence of molecules strongly depend on the temperature and the interaction between molecules. The detailed phase diagram in the interaction-temperature parameter space is derived.

The paper is organized as follows. In section II, we introduce an effective model and discuss its relevance to the cell membranes. In section III, we construct our Landau energy functional through Helmholtz free energy using a mean-field theory. In section IV, we provide the phase diagrams with detailed results. Finally, in section V we draw our conclusions with a short discussion.

\section{II. The model}

\begin{figure}
\includegraphics[width=\columnwidth]{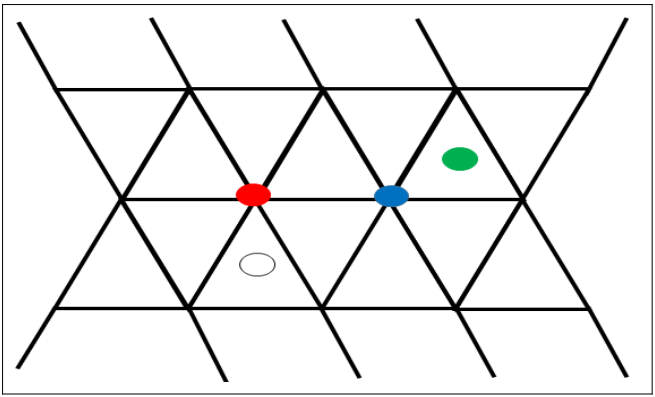}
\caption{(color online) The occupation of A-molecule (blue) and B-molecule (red) at the corners of a triangular lattice. If a given site $i$ at a triangular lattice site is occupied by a A-molecule, we assign $\sigma_i = +1$, and if a given site $j$ at a triangular lattice is occupied by a B-molecule, we assign $\sigma_j = -1$. All sites on the triangular lattice are occupied by either A-molecule of B-molecule. Notice that the centers of the triangular lattice can be connected to form a hexagonal lattice. If a center of the triangular lattice $\alpha$ (corner of a hexagonal) is occupied by a C-molecule (green), we assign $s_\alpha = +1$ and if the center of a triangular $\beta$ is empty (empty circle), then we assign $s_\beta = 0$. }\label{Ls}
\end{figure}

We assume each of the three components of the mixture can sit on a discrete lattice and interact with each other. As shown in FIG~\ref{Ls}, the saturated lipids (A) and the unsaturated lipids (B) are sitting on a two-dimensional triangular grid while the cholesterol molecule (C) is sitting at sites on a complementary hexagonal lattice. Notice that the hexagonal lattice points are located at the centers of each triangle within the triangular lattice. Assuming only the nearest-neighbor interactions, we model the interaction between molecular components using an Ising type model. In this approach, we assume that the cell is large enough to have many lattice sites. Thus, physical quantities can be calculated averaging over many cells. We define two variables $\sigma_i = \pm 1$, denoting the occupation of molecule $A$ or $B$ at site “$i$” on the triangular lattice and $s_\alpha = 1$ or $s_\alpha = 0$ denoting whether the hexagonal lattice site $\alpha$ is occupied by the molecule C or not. Then, the “energy function” or the Hamiltonian for these interacting molecules on the two-dimensional lattice grids is written in a generalized Ising model form. The fictitious lattice points for the molecular position on a cell is introduced to track the molecular positions in our mathematical calculations.

The model describing the occupation of molecules is thus given by,

\begin{eqnarray}
H = -J \sum_{\langle ij \rangle} \sigma_i \sigma_j - K \sum_{\langle i \alpha \rangle} \sigma_i (s_\alpha-\frac{1}{2}) \\ \nonumber
- L \sum_{\langle  \alpha \beta \rangle} (s_\alpha-\frac{1}{2}) (s_\beta-\frac{1}{2}) \\ \nonumber
+ M \sum_{\langle ij \alpha \beta \rangle} (\sigma_i \sigma_j) ((s_\alpha-\frac{1}{2}) (s_\beta-\frac{1}{2}),
\end{eqnarray}

\noindent where $\langle ij \rangle$ represents only the nearest neighbors on the triangular lattice and $\langle \alpha \beta \rangle$ represents only the nearest neighbors on the complimentary hexagonal lattice. In this model, all the possible interactions between the molecules on the cell are represented by the coupling between the variables $\sigma$ and $s$. The coupling strengths are represented by the molecular dependent interaction constant $J$, $K$, $L$, and $M$. For example, the first term represents the $A-B$ molecular interactions between the nearest neighbor sites $i$ and $j$ on the fictitious triangular lattice. The second terms represents the $A-C$, $A$-empty, $B-C$, and $B$-empty molecular interactions between $A$ and $B$ molecules at site $i$ on a triangular lattice, with $C$-molecules and empty site $\alpha$ on a fictitious hexagonal lattice. Similarly, the third represents the interaction between molecules on the hexagonal lattice. Finally, the fourth term represents the interaction between molecules at four neighboring sites, with sites $i$ and $j$ being on the triangular lattice and sites $\alpha$ and $\beta$ being on the hexagonal lattice. Here we assume that the interaction between molecules that are not neighboring are negligible compared to interactions with nearest neighbors.

\section{III. The Landau Energy Functional}

The model Hamiltonian represented by the Eq. (1) describes the interacting molecules, thus it is unable to solve exactly. However, the solutions can be obtained by converting it into an effective non-interacting model. In order to convert the interactive Hamiltonian in Eq. (1) into an effective non-interacting Hamiltonian, we introduce two order parameters; $\phi = \sum_i \langle \sigma_i \rangle \equiv \frac{N_A-N_B}{N_A + N_B}$ and $\Delta = \sum_\alpha \langle s_\alpha \rangle \equiv \frac{N_C}{N_C + N_E}$. Here $N_X$, with $ X= A, B, C, E$ represents the total number of A, B, C molecules and empty sites on the cell membrane, respectively. These order parameters are defined as the quantum and statistical average values of the variables, $\sigma$ and $s$. While non-zero values of $\phi$ represent the $A-B$ phase separation, non-zero values of $\Delta - 1/2$ represent the phase separation of C molecules and empty sites. By replacing the interaction variables with these order parameters, the internal energy $E = \langle H \rangle$ of the system can be written as,

\begin{eqnarray}
E = -J z \phi^2 - Kz \phi (\Delta-\frac{1}{2})-Lz (\Delta-\frac{1}{2})^2 \nonumber \\
+ Mz \phi^2 (\Delta-\frac{1}{2})^2,
\end{eqnarray}

\noindent where $z = 6$ is the number of nearest neighbors. The entropy of the system $S = -k_B \ln[G_{ab} G_{ce}]$ is then evaluated using,

\begin{eqnarray}
G_{ab} = \frac{N_{tt} !}{N_A ! N_B !} \\ \nonumber
G_{ce} = \frac{N_{th} !}{N_C ! N_E !},
\end{eqnarray}

\noindent where $G_{ab}$ and $G_{ce}$ represent all the possible arrangements of molecules and empty sites on triangular and hexagonal lattices. The entropy is then evaluated to be,

\begin{eqnarray}
S = -\frac{1-\phi}{2} \ln \biggr[\frac{1-\phi}{2} \biggr] -\frac{1+\phi}{2} \ln \biggr[\frac{1+\phi}{2}\biggr] \\ \nonumber
- \Delta \ln \Delta – (1-\Delta) \ln[1-\Delta].
\end{eqnarray}

\begin{figure}
\includegraphics[width=\columnwidth]{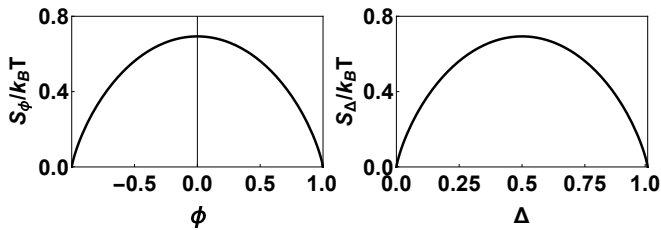}
\caption{The dimensionless entropy $S/k_BT = S_\phi/k_BT + S_\Delta/k_BT$ as a function of order parameters $\phi$ and $\Delta$. Notice that the entropies maximize at $\phi =0$ and $\Delta = 0.5$.}\label{ENP}
\end{figure}

\noindent The entropy in thermal units $k_BT$ is plotted in FIG.~\ref{ENP}, where $T$ is the temperature in Kelvins and $k_B$ is the Boltzmann constant. Notice the entropy has a maximum at $\phi = 0$ and $\Delta = 0.5 $. These are entropy dominated high-temperature values of order parameters. In the following, we expand the Helmholtz free energy ($F = E-TS$) around these high-temperature values of order parameters up to the fourth order to obtain the Landau energy functional. The Landau energy functional is the order by order expansion of the Helmholtz free energy up to the fourth order in order parameters.

\noindent Defining a new scaled order parameter $\alpha = \Delta -1/2$, the Landau energy functional can be written as,

\begin{eqnarray}
\frac{F}{k_BT} = \frac{1}{2} R_1 \phi^2 + \frac{1}{2} R_2 \alpha^2 + \frac{1}{4} Q_1 \phi^4 + \frac{1}{4} Q_2 \alpha^4 \frac{1}{2} P \phi^2 \alpha^2,
\end{eqnarray}

\noindent where we introduced four new dimensionless parameters; $R_1 = 1- \frac{2Jz}{k_BT}$, $R_2 = 1- \frac{2Lz}{k_BT}$, $Q_1 = \frac{1}{3}$, $Q_2 = \frac{16}{3}$, and $P = \frac{2Mz}{k_BT}$, all dependent on the interaction parameters and temperature. The molecular organization on the cell membrane is then determined by the equilibrium values of order parameters $\phi$ and $\alpha$. Notice that $\alpha$ which is related to the order parameter $\Delta$ is introduced for convenience.

\section{IV. Results}

\begin{figure}
\includegraphics[width=\columnwidth]{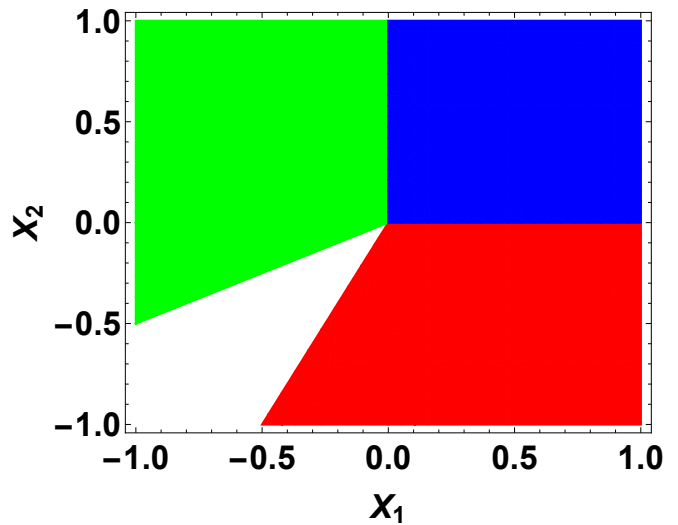}
\caption{(color online) The phase diagram of the three-component mixture at $\lambda = +0.5$. The blue region represents the homogeneous mixture, where all three components co-exist in the same region of space. While the red region represents the C-molecule-empty mixed phase, the green region represents the molecular A-B mixed phase. In the red region, A and B molecules are separated. The green region, C-molecules and empty sites are separated. The white region represents the completely phase separated state. See the text for details.}\label{PD2}
\end{figure}

\begin{figure}
\includegraphics[width=\columnwidth]{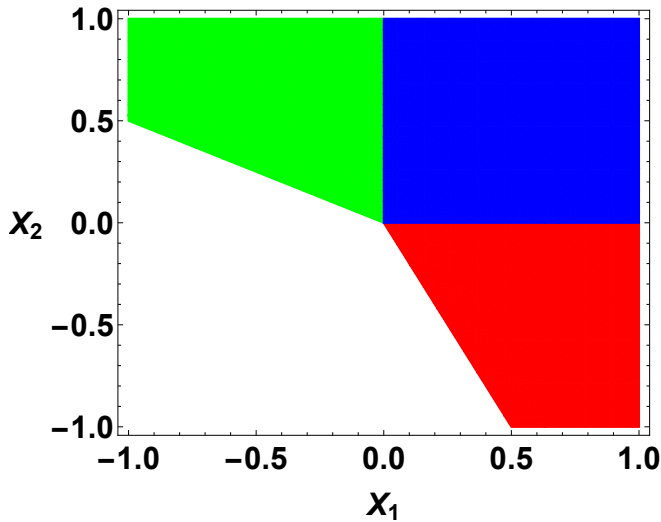}
\caption{(color online) The phase diagram of the three-component mixture at $\lambda = -0.5$. The color code is the same as that of FIG~.\ref{PD2}. }\label{PD3}
\end{figure}

The equilibrium values of the order parameters depend on the interaction parameters and the temperature. For a given set of parameters, the molecules $A-B$ are phase separated on the cell when the order parameter $\phi \neq 0$. In other words, when $\phi = 0$ the molecules $A$ and $B$ are mixed on the triangular lattice. Similarly, when the order parameter $\alpha = 0 (\Delta = 1/2)$, the C molecules and the empty sites are mixed on the hexagonal lattice. As a result, the homogeneous phase is determined by the simultaneous conditions $\phi = 0$ and $\alpha = 0$ for a given set of interaction parameters and temperatures. If these simultaneous conditions are not met, then we have three different types of inhomogeneous phases depending on the values of $\phi$ and $\alpha$. By minimizing the Landau energy functional with respect to the order parameters, we can construct the phase diagram of the system in interaction-temperature parameter space. By taking the derivative of the energy functional with respect to $\phi$ and $\alpha$ and then equating them to zero for minima, we analytically solve the minimization equations for equilibrium values for order parameters using three scaled parameters $X_1 = R_1/\sqrt{Q_1}$, $X_2 = R_2/\sqrt{Q_2}$, and $\lambda = P/\sqrt{Q_1 Q_2}$. Notice that during our functional derivatives, the five temperature dependent dimensionless parameters, $R_1$, $R_2$, $Q_1$, $Q_2$, and $P$ are replaced in the expense of \emph{only} three temperature dependent dimensionless parameters, for mathematical simplicity. The results are summarized in the phase diagrams shown in FIG~\ref{PD2} and FIG~\ref{PD3}.

For both $X_1 > 0$ and $X_2 >0$, we find equilibrium values of the order parameters $\alpha = \phi = 0$. These values of order parameters represent the mixed phase or the homogeneous phase of the cell membrane where all molecules $A$, $B$, $C$, and empty sites on fictitious hexagonal lattice are mixed together. In the phase diagrams shown in FIG~\ref{PD2} and FIG~\ref{PD3}, this phase is shown in blue in our reduced parameter space. For both $X_1 < 0$ and $X_2 < 0$, or when one of them is negative, we find completely phase separation or partially phase separation of phases. These depend on the interaction and temperature dependent parameters  $X_1$, $X_2$, and $\lambda$. AS examples, we construct phase diagrams for two representative values of the parameter $\lambda = +0.5$ and $\lambda = +0.5$ in FIG.~\ref{PD2} and FIG.~\ref{PD3}, respectively. The green region where $X_1 < 0$ represents the molecular $A-B$ phase separated phase with mixed $C$ and empty sites, where the order parameters $\phi \neq 0$ and $\alpha = 0$. On the other hand, the red region represents the $C$ molecules and empty sites separated phase with mixed $A-B$ molecular phase, where $\phi = 0$ and $\alpha \neq 0$. The white region represents the phase separation where the molecules $A$ and $B$ separated, but coexist with either the empty sites or $C$ molecules. Notice that the phase diagram is very sensitive to the parameter $\lambda$. When $\lambda = 0$, the phase diagram is symmetric in $X_1-X_2$ parameter space, where the four phases exist in four quadrants symmetrically. On the other hand, when $\lambda = 1$, the green and red region merge at $(X_1, X2) = (-1.0, -1.0)$ while the white region disappears completely. Otherwise, the qualitative features of the phase diagram for other values of $\lambda$ remain the same with the homogeneous phase in the first quadrant in $X_1-X_2$ parameter space and sharing the other three phases in the remaining region in the phase diagram.

\section{V. Discussion and Conclusions}

In this work, we used a thermodynamical approach to study the self-organization of a three-component molecular mixture on a two-dimensional membrane. By treating molecular interactions between localized molecules as spin variables we study the formation of the homogeneous phases and various phase separated states. We argued that the different phases emerge due to the competition between entropy and the energy originated from the collective behaviour of molecules. We constructed a detailed phase diagram by introducing two order parameters through a mean field theory to specify the different composition of the mixture. Our phase diagram was deduced by minimizing the Landau-type energy functional derived from the Helmholtz free energy. We find that the rich phase diagram is very sensitive to the microscopic interaction parameters between molecules and the temperature.

In this work, we neglected the structure of the molecules thus enabling our ability to model the cell membrane as a two-dimensional surface. Although, our qualitative phase diagram is accurate for a structureless three-component mixture, the molecular structure and the orientation must be included for more realistic membranes. These can be included, for example, by assuming that phospholipids are “rod-like” molecules which maintain their conformation through dipolar- dipolar interactions. This dipolar interaction gives negative interaction energies for the parallel oriented rod like molecules on two different planes. Thus, the negative interaction forces two rod-like molecules to form bound states and these bound states are responsible for the formation of bi-layers.

\section{VI. Acknowledgments}

The authors acknowledge the support of Augusta University. This research was supported in part by the Augusta University Provost's office, and the Translational Research Program of the Department of Medicine, Medical College of Georgia at Augusta University. The authors thank Peighton Bolt for reading and providing critical comments on the manuscript.

\end{document}